\documentclass[a4paper,twoside,fleqn,12pt]{article}

\usepackage[final]{graphicx}
\usepackage{graphics}
\usepackage[mathscr]{eucal}
\usepackage{euler}
\usepackage{amsmath}
\usepackage{amssymb}
\usepackage{subeqnarray}
\usepackage{graphicx} 
\usepackage{epsfig}

\usepackage[english]{babel}

\def\booktitle{APM Proceedings}

\def\SelectLanguage{\selectlanguage{english}}

\usepackage{fancyhdr}

\usepackage{Tabbing}

\def\email#1{\date{\tt#1}}
\def\address#1{\par\noindent#1\smallskip}

     \renewcommand{\sectionmark}[1]%
     {}{}
     \renewcommand{\subsectionmark}[1]%
     {}
\makeatletter
\renewenvironment{thebibliography}[1]
     {\section*{\refname}%
      \list{\@biblabel{\@arabic\c@enumiv}}%
           {\settowidth\labelwidth{\@biblabel{#1}}%
            \leftmargin\labelwidth
            \advance\leftmargin\labelsep
            \@openbib@code
            \usecounter{enumiv}%
            \let\p@enumiv\@empty
            \renewcommand\theenumiv{\@arabic\c@enumiv}}%
      \sloppy
      \clubpenalty4000
      \@clubpenalty \clubpenalty
      \widowpenalty4000%
      \sfcode`\.\@m}
     {\def\@noitemerr
       {\@latex@warning{Empty `thebibliography' environment}}%
      \endlist}
\makeatother

\def\Title#1{%
    \SelectLanguage
    \title{~\\ [20mm]{#1}}
    \date{}
    \maketitle
    \thispagestyle{fancy} 
    \markboth{\booktitle}{\shorttitle} 
    \vspace{-15mm} 
}

\def\papertitle{Fractional Lattice Dynamics: Nonlocal constitutive behavior generated by power law matrix functions and their fractional continuum limit kernels}
\def\paperauthor{\begin{center} T.M.~Michelitsch, \and B.A.~Collet, \and A.P.~Riascos,
\and A.F.~Nowakowski, \and F.C.G.A.~Nicolleau \end{center}}
\def\paperemail{\hspace{2cm} michel@lmm.jussieu.fr 
\newline\newline  To appear in APM 2016 proceedings http://www.apm-conf.spb.ru/
}

\def\titlewithbreakes{\papertitle}
\def\shorttitle{Fractional Lattice Dynamics}

\begin{document}

\begin{center}
\Title{\titlewithbreakes}
\author{\paperauthor}
\email{\paperemail}
\end{center}

\begin{abstract}
We introduce positive elastic potentials in the harmonic approximation leading by Hamilton's variational principle to fractional Laplacian matrices having the forms of power law matrix functions of the simple local Born
von Karman Laplacian. The fractional Laplacian matrices are well defined on periodic and infinite lattices in $n=1,2,3,..$ dimensions. The present approach generalizes the central symmetric second difference
operator (Born von Karman Laplacian) to its fractional central symmetric counterpart (Fractional Laplacian matrix).

For non-integer powers of the Born von Karman Laplacian, the fractional Laplacian matrix is nondiagonal with nonzero matrix elements everywhere, corresponding to nonlocal behavior: For large lattices the matrix elements far from the diagonal expose power law asymptotics leading to continuum limit kernels of 
Riesz fractional derivative type. We present explicit results for the fractional Laplacian matrix in 1D for finite periodic and infinite linear chains and their Riesz fractional derivative continuum limit kernels.
The approach recovers for $\alpha=2$ the well known classical Born von Karman linear chain (1D lattice) with local next neighbor springs
leading in the well known continuum limit of classic local standard elasticity, and for other integer powers to gradient elasticity.
We also present a generalization of the fractional Laplacian matrix to n-dimensional cubic periodic (nD tori) and infinite lattices. 
For the infinite nD lattice we deduce
a convenient integral representation.
We demonstrate that our fractional lattice approach 
is a powerful tool to generate physically admissible nonlocal  lattice material models and their continuum representations.

\end{abstract}

\section{Introduction}
Fractional calculus has turned out to be a powerful analytical tool in various disciplines: It has been recognized that
especially in more recently emerging fields dealing with complex, chaotic, turbulent, critical, fractal and anomalous transport phenomena, problems can appropriately be described by equations
which involve fractional operators. A broad overview on applications of the fractional approach can be found in the review articles of Metzler and Klafter \cite{metzler,metzler2014}.

There exist various definitions (Riemann, Liouville, Caputo, Gr\"unwald-Letnikow, Marchaud, Weyl, Riesz, Feller, and others) for fractional derivatives and integrals, e.g. \cite{hilfer-2008,metzler,samko,samko2003,podlubny} among many others. 
This diversity of definitions is due to the fact that fractional operators take different kernel representations in different function spaces which is a consequence of the nonlocal character of fractional kernels.

The present paper is organized as follows. In the first part of the paper
we deduce from ``fractional harmonic lattice potentials" on the cyclically closed linear chain a discrete fractional Laplacian matrix.
We do so by applying our recent approach to generate
nonlocal lattice models by matrix functions where the generator operator is the discrete centered Born von
Karman Laplacian \cite{michel-collet}. First we obtain the discrete fractional Laplacian
in explicit form for the infinite chain for particle numbers $N\rightarrow \infty$,
being in accordance with the fractional centered difference models of Ortiguiera \cite{riesz2}
and Zoia et al. \cite{Zoia2007}.
Utilizing the discrete infinite chain fractional Laplacian matrix we construct an explicit representation for the {\it fractional Laplacian matrix on the $N$-periodic finite 1D lattice} where the particle number
$N$ can be arbitrary not necessarily large.
Then we analyse continuum limits of the discrete fractional model:
The infinite space continuum limit of the fractional Laplacian matrix yields the well known infinite space kernel
of the standard fractional Laplacian. The periodic string continuum
limit yields an explicit representation for the kernel of the fractional Laplacian (Riesz fractional derivative) which fulfills periodic boundary conditions and is defined on the finite $L$-periodic string.  

In the second part of the paper we suggest an extension of the fractional approach on nD periodic and infinite lattices. We deduce an integral representation for fractional Laplacian on the infinite nD lattice and proof that
as asymptotic representation the well known Riesz fractional derivative of the nD infinite space is emerging. 
More detailed derivations of some of the results of the present paper can be found in recent articles
\cite{michelJphysA,michelchaos}. All these results are fully equivalent and can also be deduced by employing the more general 
approach of Riascos and Mateos for fractional diffusion problems on networks \cite{riascos-fracdyn,riascos-fracdiff}, and see also the references therein.

\section{Fractional Laplacian matrix on the finite periodic 1D lattice}

We consider a periodic, cyclically closed linear chain (1D periodic lattice or ring) with equidistant lattice points $p=0,..,N-1$ consisting of $N$ identical particles having all the same mass $\mu$. 
Each mass point $p$ has equilibrium position at $0\leq x_p=ph < L=Nh $ ($p=0,..,N-1$) where $L$ denotes the length of the chain and $h$ the interparticle distance (lattice constant). 
Further we impose periodicity (cyclic closure of the chain). For convenience of our demonstration we introduce the unitary shift operator $D$ defined by $ Du_p= u_{p+1}$ and its adjoint $D^{\dagger}= D^{-1}$ with $D^{\dagger}u_p = u_{p-1}$.
We employ periodic boundary conditions (cyclic closure of the chain) $u_p=u_{p+sN}$ ($s\in {\bf Z}$) and equivalently, cyclic index convention $p \rightarrow p \,\,\, mod\, (N)  \in \{0,1,..,N-1\}$.
Any elastic potential in the {\it harmonic approximation} defined on the 1D periodic lattice can be written in the representation \cite{michel-collet}
\begin{equation}
\label{compfo}
V_f = \frac{\mu}{2}\sum_{p=0}^{N-1}u_p^*f(2{\hat 1}-D-D^{\dagger})u_p = -\frac{1}{2} \sum_{p=0}^{N-1}\sum_{q=0}^{N-1} u_q^*\Delta_f(|p-q|)u_p ,
\end{equation}
where $\Delta_f(|p-q|) = -\mu f_{|p-q|}$ indicates the (negative-semidefinite) Laplacian $N\times N$-matrix, ${\hat 1}$ the identity matrix, and $f$ we refer to as the characteristic function: 
Physically admissible, elastically stable and translational invariant positive elastic potentials require for the 1D periodic lattice (cyclic ring) that the
characteristic function $f$ which is defined as a scalar function to have the following properties $0<f(\lambda) < \infty$ for $0<\lambda \leq 4$ (elastic stability) and $f(\lambda=0)=0$ (translational invariance, zero elastic energy for uniform translations of the lattice). For the approach to be developed we propose the characteristic function to assume power law
form
\begin{equation}
\label{powerlaw}
f^{(\alpha)}(\lambda) = \Omega_{\alpha}^2\lambda^{\frac{\alpha}{2}} ,\hspace{2cm} \alpha >0 ,
\end{equation}
which fulfills for $\alpha >0$ and $\Omega_{\alpha}^2 >0$ the above required good properties for the characteristic function. $\Omega_{\alpha}$ denotes a dimensional constant of physical dimension $sec^{-1}$. 
Note that $2{\hat 1}-D-D^{\dagger}$
is the central symmetric second difference operator which is defined by $(2{\hat 1}-D-D^{\dagger})u_p=2u_p-u_{p+1}-u_{p-1}$. The matrix function $f(2{\hat 1}-D-D^{\dagger})$ is in general a  self-adjoint (symmetric)
positive semidefinite $N\times N$-matrix function of the simple $N\times N$ generator matrix $[2{\hat 1}-D-D^{\dagger}]_{pq}=2\delta_{pq}-\delta_{p+1,q}-\delta_{p-1,q}$. It can be easily seen that $f(2{\hat 1}-D-D^{\dagger})$ 
has T\"oplitz structure, i.e. its additional symmetry consists in the form
$f_{pq}=f_{qp}=f_{|p-q|}$, $p,q=0,..N-1$ giving the fractional generalization of the Born von Karman centered difference operator. The fractional elastic potential has then with (\ref{compfo}) the representation
\begin{equation}
\label{Valpha}
V_{\alpha} = \frac{\mu\Omega_{\alpha}^2}{2} \sum_{p=0}^{N-1}u_p^*(2-D-D^{\dagger})^{\frac{\alpha}{2}}u_p =
\frac{\mu}{2} \sum_{p=0}^{N-1}\sum_{q=0}^{N-1}u_q^*f^{(\alpha)}_{|p-q|}u_p ,
\end{equation}
with the matrix elements $f^{(\alpha)}_{|p-q|} =\Omega_{\alpha}^2[(2\,{\hat 1}-D-D^{\dagger})^{\frac{\alpha}{2}}]_{|p-q|}$
of the {\it fractional characteristic matrix function}. In full analogy to the negative semidefinite continuous Laplacian (second derivative operator) we define here the fractional Laplacian matrix as the negative semidefinite matrix defined through Hamilton's variational principle\footnote{The sign convention differes in many references, so e.g. in \cite{riascos-fracdyn,riascos-fracdiff} the fractional Laplacian matrix is defined positive semidefinite corresponding to the definition of the characteristic fractional operator $(2{\hat 1}-D-D^{\dagger})^{\frac{\alpha}{2}}$. }
\begin{equation}
\label{fractlap}
\Delta_{\alpha} u_p = -\frac{\partial}{\partial u_p}V_{\alpha} , \hspace{2cm}
\Delta_{\alpha} = -\mu\Omega_{\alpha}^2 (2-D-D^{\dagger})^{\frac{\alpha}{2}} .
\end{equation}
It turns out that the fractional Laplacian matrix for non-integer $\frac{\alpha}{2}$ has nonzero matrix elements everywhere with power law asymptotics $\sim |p-q|^{-\alpha-1}$ for $|p-q| >>1$ sufficiently large
corresponding to nonlocal constitutive behavior. For the finite 1D periodic lattice the matrix elements of the fractional Laplacian matrix (\ref{fractlap}) will be evaluated explicitly which we shall do in the subsequent paragraph.
\newline\newline
{\bf Infinite 1D lattice} \newline\newline
This can be done using its spectral representation which assumes in the limiting case of an infinite lattice ($N\rightarrow \infty$), see e.g. \cite{michelJphysA,michelchaos,riascos-fracdyn}
\begin{equation}
\label{fractlattice}
\begin{array}{l}
\displaystyle f^{(\alpha)}_{|p-q|}=\Omega_{\alpha}^2(2-D-D^{\dagger})^{\frac{\alpha}{2}}_{|p-q|} ,\hspace{1cm} p,q \in {\bf Z}_0  ,\\  \\
\displaystyle f^{(\alpha)}_{|p|} ,
=\frac{\Omega_{\alpha}^2}{2\pi}\int_{-\pi}^{\pi}e^{i\kappa p}\left(4\sin^2{\frac{\kappa}{2}}\right)^{\frac{\alpha}{2}} {\rm d}\kappa .
\end{array}
\end{equation}
This expression can be obtained in explicit form \cite{michelJphysA,michelchaos,Zoia2007}
\begin{equation}
\label{matrixelei}
f^{(\alpha)}(|p|) = \Omega_{\alpha}^2\,\frac{\alpha!}{\frac{\alpha}{2}!(\frac{\alpha}{2}+|p|)!}(-1)^p\prod_{s=0}^{|p|-1}(\frac{\alpha}{2}-s) = 
\Omega_{\alpha}^2 \,(-1)^p\, \frac{\alpha!}{(\frac{\alpha}{2}-p)!(\frac{\alpha}{2}+p)!} ,
\end{equation}
where we introduced the generalized factorial function $\beta ! = \Gamma(\beta+1)$. In view of (\ref{matrixelei}) we observe that for noninteger $\frac{\alpha}{2}$ any matrix element $f^{(\alpha)}(|p-q|) \neq 0$ is non-vanishing indicating the nonlocality of the harmonic fractional interparticle interaction (\ref{fractlap}). 
For $\frac{\alpha}{2} = m \in {\bf N}$ the matrix elements (\ref{matrixelei}) take the values of the standard binomial coefficients.
(\ref{fractlattice})$_2$ can be read as the Fourier coeffcients of the infinite Fourier series
\begin{equation}
\label{Fouser}
\omega_{\alpha}^2(\kappa) = \Omega_{\alpha}^2\left(4\sin^2{\frac{\kappa}{2}}\right)^{\frac{\alpha}{2}} = 
\Omega_{\alpha}^2\left(2-e^{i\kappa }-e^{-i\kappa }\right)^{\frac{\alpha}{2}} = \sum_{p=-\infty}^{\infty}f^{(\alpha)}_{|p|}e^{ip \kappa} .
\end{equation}
This relation indicates the fractional dispersion relation of the infinite lattice leading to the remarkable
relation which holds {\it only} for complex numbers on the unit circle $z=e^{i\kappa }$, namely
\begin{equation}
\label{Fouserb}
\left(2-z-\frac{1}{z}\right)^{\frac{\alpha}{2}} = \sum_{p=-\infty}^{\infty}(-1)^p\, \frac{\alpha!}{(\frac{\alpha}{2}-p)!(\frac{\alpha}{2}+p)!}z^{p} ,
\hspace{2cm} |z|=1 .
\end{equation}
This Laurent series converges nowhere except on the unit circle $|z|=1$.
For instance the zero eigenvalue $\omega_{\alpha}^2(\kappa=0)=0$ which corresponds to translational invariance (zero elastic energy for uniform translations) 
is obtained by putting $z=1$ in (\ref{Fouserb}).
For integer $\frac{\alpha}{2}=m \in {\bf N}$ (\ref{matrixelei}) takes the form of the standard binomial coefficients
and the series (\ref{Fouser}), (\ref{Fouserb}) then take the representations of standard binomial series of
$\left(2-z-\frac{1}{z}\right)^{\frac{\alpha}{2}}=(-1)^m(\sqrt{z}-\frac{1}{\sqrt{z}})^{2m}$ breaking at $|p|=m$ corresponding to zero values for the matrix elements for (\ref{matrixelei}) for $|p|>m$.
We further observe for noninteger $\frac{\alpha}{2} \notin {\bf N}$ the {\it power law asymptotics} for $|p|>>1$ which can be obtained by utilizing Stirling's asymptotic formula for the $\Gamma$-function \cite{michelJphysA,michelchaos}
\begin{equation}
\label{asymp}
f^{(\alpha)}_{|p|>>1} \rightarrow -\Omega_{\alpha}^2\,\,\frac{\alpha!}{\pi}\sin{(\frac{\alpha\pi}{2})} \,\, p^{-\alpha-1} .
\end{equation}
The asymptotic power law (scale free) characteristics of the fractional Laplacian matrix $\Delta_{pq} \sim |p-q|^{-\alpha-1}$ is the essential property which gives rise to many 
`anomalous phenomena' such as in `fractional diffusion' problems on networks such as the emergence of L\'{e}vy flights \cite{riascos-fracdyn,riascos-fracdiff} (and references therein). The fractional continuum limit kernels are discussed in the subsequent section.
The expressions (\ref{fractlattice})-(\ref{asymp}) hold for the infinite 1D lattice corresponding to $N\rightarrow\infty$.
As everything in nature is limited we shall consider now the fractional Laplacian matrix for a finite periodic lattice
where the particle number $N$ is arbitrary and not necessarily large.
\newline\newline
{\bf 1D finite periodic lattice - ring}
\newline\newline
It is only a small step to construct the finite lattice Laplacian matrix in terms of infinite lattice Laplacian matrix.
We can perform this step by the following consideration: Let $-\mu f^{(\infty)}_{|p-q|}$ the Laplacian matrix of the infinite lattice, and $\omega^2(\kappa)$ the continuous dispersion relation of the infinite lattice matrix $f_{|p-q|}$ obeying the eigenvalue relation

\begin{equation}
\label{eigval}
\sum_{q=-\infty}^{\infty}f^{(\infty)}_{|p-q|}e^{iq\kappa} = \omega^2(\kappa)e^{ip\kappa} ,\hspace{2cm} 0\leq \kappa < 2\pi .
\end{equation}
This relation holds identically in the entire principal interval $0\leq \kappa < 2\pi$ and is $2\pi$-periodic in the $\kappa$-space. Let us now choose 
$\kappa=\kappa_{\ell}=\frac{2\pi}{N}\ell$ with $\ell=0,..,N-1$ being the Bloch wave number of the
{\it finite} periodic lattice of $N$ lattice points where $N$ is not necessarily large. Since the Bloch wave numbers of the chain are discrete points within the interval $0\leq \kappa_{\ell}<2\pi$, then relation (\ref{eigval}) 
holds as well for these $N$ $\kappa$-points, namely
\cite{michelJphysA,michelchaos}\footnote{where $p=0$ in (\ref{eigval}) has been put to zero.}
\begin{equation}
\label{eigvalblock}
\begin{array}{l}
\displaystyle \sum_{p=-\infty}^{\infty}f^{(\infty)}_{|q|}e^{iq\kappa_{\ell}} = 
\omega^2(\kappa_{\ell}) ,\hspace{2cm} 0\leq \kappa_{\ell}=\frac{2\pi}{N}\ell < 2\pi , \nonumber \\ \nonumber\\
\displaystyle \sum_{p=0}^{N-1}\sum_{s=-\infty}^{\infty}f^{(\infty)}_{|p+sN|} e^{i(p+sN)\kappa_{\ell}} = \sum_{p=0}^{N-1}e^{ip\kappa_{\ell}}\sum_{s=-\infty}^{\infty}f^{(\infty)}_{|p+sN|} =
\sum_{p=0}^{N-1}e^{ip\kappa_{\ell}} f^{finite}_{|p|} =  \omega^2(\kappa_{\ell}) .
\end{array}
\end{equation}
In the second relation the $N$-periodicity of the finite lattice Bloch eigenvector $e^{i(p+sN)\kappa_{\ell}}= e^{ip\kappa_{\ell}}$  has been used. 
The last relation can be read as the eigenvalue relation for the $N$-periodic lattice matrix of T\"oplitz structure
\begin{equation}
\label{identiperiodic}
f^{finite}_{|p-q|} =
\sum_{s=-\infty}^{\infty}f^{(\infty)}_{|p-q+sN|} = f^{(\infty)}_{|p-q|}+ \sum_{s=1}^{\infty}(f^{(\infty)}_{|p-q+sN|}+f^{(\infty)}_{|p-q-sN|}) .
\end{equation}
It follows that in the limiting case $N\rightarrow\infty$ the finite lattice matrix (\ref{identiperiodic}) recovers the infinite lattice matrix $f^{finite} \rightarrow f^{(\infty)}$. From (\ref{identiperiodic}) we read of for the fractional lattice Laplacian of the finite periodic 1D lattice

\begin{equation}
\label{finitefraclap}
\Delta_{\alpha,N}(|p|) = -\mu f^{(\alpha,finite)}_{|p|} ,\hspace{1cm} 0\leq p \leq N-1 
\end{equation}
with
\begin{equation}
\label{finitecharmat}
\begin{array}{l}
f^{(\alpha,finite)}_{|p|} =  \Omega_{\alpha}^2\displaystyle \frac{(-1)^p\alpha!}{(\frac{\alpha}{2}-p)!(\frac{\alpha}{2}+p)!} + \Omega_{\alpha}^2 \sum_{s=1}^{\infty}(-1)^{p+Ns}\alpha !
\left(  \frac{1}{(\frac{\alpha}{2}-p-sN)!(\frac{\alpha}{2}+p+sN)!} \right. \\ \\ 
\left. \hspace{2cm} \displaystyle + \frac{1}{(\frac{\alpha}{2}-p+sN)!(\frac{\alpha}{2}+p-sN)!} \right) .
\end{array}
\end{equation}
We observe $N$-periodicity of (\ref{finitecharmat}) and furthermore the necessary property that in the limit of infinite chain $N\rightarrow \infty$, (\ref{finitecharmat}) recovers the infinite lattice expression
of eq. (\ref{matrixelei}).

\section{Fractional continuum limit kernels}

In this section we investigate the interlink between the lattice fractional approach introduced above and continuum fractional derivatives. To this end we introduce the following hypotheses which are to be observed when performing
continuum limits. Following \cite{michel-collet} we require in the continuum limit that extensive physical quantities, i.e. quantities
which scale with the length of the 1D system, such as the total mass $N\mu=M$ and the total elastic
energy of the chain remain finite when its length $L$ is kept finite\footnote{In the case of infinite string $L\rightarrow \infty$ we require the
mass per unit length and elastic energy per unit length to remain finite.}, i.e. neither vanish nor diverge. Let $L=Nh$ be the length of the chain and $h$ the lattice constant 
(distance between two neighbor atoms or lattice points).

\noindent We can define two kinds of continuum limits: \newline\noindent  (i) The {\it periodic string continuum limit} where the length of the chain
$L=Nh$ is kept finite and $h\rightarrow 0$ (i.e. $N(h)= L h^{-1} \rightarrow \infty$). \newline\noindent (ii) The {\it infinite space continuum limit} where $h\rightarrow 0$, 
however, the length of the chain tends to infinity $N(h)h=L(h) \rightarrow \infty$\footnote{which can be realized for instance by chosing by $N(h)\sim h^{-\delta}$ where $\delta > 1$.}. The kernels of the infinite space limit can be recovered from those of
the periodic string limit by letting $L\rightarrow \infty$.
From the finiteness of
the total mass of the chain, it follows that the particle mass $\mu=\frac{M}{N}=\frac{M}{L} h = \rho_0 h$ scales as $\sim h$.
 Then by employing expression (\ref{Valpha}) for the fractional elastic potential,
the total continuum limit elastic energy ${\tilde V}_{\alpha}$ can be defined by
\begin{equation}
\label{elasten}
{\tilde V}_{\alpha} = 
\lim_{h\rightarrow 0+} V_{\alpha} = 
\frac{\mu\Omega_{\alpha}^2}{2}\sum_{p=0}^{N-1} u^*(x_p)\left(-4\sinh^2{\frac{h}{2}\frac{d}{dx}}\right)^{\frac{\alpha}{2}}u(x_p) .
\end{equation}
Putting $D=e^{h\frac{d}{dx}}$ ($ph=x_p \rightarrow x$) and accounting for $2-D(h)-D(-h) =-4\sinh^2{\frac{h}{2}\frac{d}{dx}} \approx -h^2\frac{d^2}{dx^2} + O(h^4)$ we get
\begin{equation}
\label{fraclap}
\lim_{h\rightarrow 0} \left(-4\sinh^2{\frac{h}{2}\frac{d}{dx}}\right)^{\frac{\alpha}{2}} = h^{\alpha} (-\frac{d^2}{dx^2})^{\frac{\alpha}{2}} .
\end{equation}

The formal relation (\ref{fraclap}) shows that the continuum limit kernels to be deduced in explicit forms have the interpretation of the {\it Fractional Laplacian} or also in the literature referred to as {\it Riesz Fractional Derivative}.
To maintain finiteness of the elastic energy in the continuum limit $h\rightarrow 0$ the following scaling relations for the characteristic model constants, the mass $\mu$ and the frequency $\Omega_{\alpha}$ are required
\cite{michelJphysA,michelchaos}
\begin{equation}
\label{scaling}
\Omega_{\alpha}^2(h) =A_{\alpha} h^{-\alpha} , \hspace{1cm} \mu(h) = \rho_0 h ,\hspace{1cm} A_{\alpha}, \rho_0 >0
\end{equation}
where $\rho_0$ denotes the mass density with dimension $g \times cm^{-1}$ and $A_{\alpha}$ denotes a positive dimensional constant
of dimension $ sec^{-2}\times cm^{\alpha}$, where the new constants $\rho_0, A_{\alpha}$ are independent of $h$. Note that the dimensional
constant $A_{\alpha}$ is only defined up to a non-dimensional positive
scaling factor as its absolute value does not matter due to the scale-freeness of the power law.
We obtain then as continuum limit of the elastic energy by taking into account $\sum_{p=0}^{N-1}h G(x_p) \rightarrow \int_0^L G(x){\rm d}x$ and $h\rightarrow dx$, 
$x_p\rightarrow x$,
\begin{equation}
\label{contilimielasten}
\begin{array}{l}
\displaystyle
{\tilde V}_{\alpha} = \lim_{h\rightarrow 0} \frac{\mu(h)}{2}
\sum_{q=0}^{N-1}\sum_{p=0}^{N-1}u_q^* f^{(\alpha)}_N(|p-q|)u_p \nonumber \\ \nonumber \\
\displaystyle
{\tilde V}_{\alpha} = \frac{\rho_0 A_{\alpha}}{2}
\int_0^Lu^*(x)\left(-\frac{d^2}{dx^2}\right)^{\frac{\alpha}{2}}u(x)\,{\rm d}x
=: -\frac{1}{2} \int_0^L\int_0^Lu^*(x'){\tilde \Delta}_{\alpha}(|x-x'|)u(x){\rm d}x{\rm d}x' .
\end{array}
\end{equation}
The continuum limit Laplacian kernel ${\tilde \Delta}_{\alpha}(|x-x'|)$ can then formally be represented by the distributional kernel 
representation in the spirit of generalized functions \cite{gelfand}
\begin{equation}
\label{contilimlap}
{\tilde \Delta}_{\alpha,L}(|x-x'|) = -\rho_0A_{\alpha}\left(-\frac{d^2}{dx^2}\right)^{\frac{\alpha}{2}}\delta_L(x-x') .
\end{equation}

The last relation contains the distributional representation of the fractional Laplacian and is obtained for the infinite space limit (ii) in explicit form as 
\cite{michelJphysA,michelchaos}
\begin{equation}
 \label{inffyla}
 {\cal K}_{\infty}^{(\alpha)}(x)= -\left(-\frac{d^2}{dx^2}\right)^{\frac{\alpha}{2}}\delta_L(x-x')=-\frac{\alpha !}{\pi} \lim_{\epsilon\rightarrow 0+} \Re
\frac{i^{\alpha+1}}{(x+i\epsilon)^{\alpha+1}} ,
\end{equation}
being defined `under the integral' which yields for noninteger $\frac{\alpha}{2}\notin {\bf N}$ for $x\neq 0$ the well known Riesz fractional derivative kernel 
of the infinite space ${\cal K}_{\infty}^{(\alpha)}(x)= \frac{\alpha ! \sin{(\frac{\alpha\pi}{2})}}{\pi}\frac{1}{|x|^{\alpha+1}}$ with a characteristic 
$|x|^{-\alpha-1}$ power law nonlocality reflecting the asymptotic
power law behavior (\ref{asymp}) of (\ref{matrixelei}) for sufficiently large $|p|>>1$.

\subsection{(i) Periodic string continuum limit}
The continuum procedure of $L$-periodic string where $L$ is kept finite is then obtained as \cite{michelJphysA,michelchaos}\footnote{where $\Re(..)$ denotes the real part of a quantity $(..)$}

\begin{equation}
\label{perfrac}
\begin{array}{l}
\displaystyle -\left(-\frac{d^2}{dx^2}\right)^{\frac{\alpha}{2}}\delta_L(x) = 
K_L^{(\alpha)}(|x|) = \frac{\alpha ! \sin{(\frac{\alpha\pi}{2})}}{\pi} \sum_{n=-\infty}^{\infty}\frac{1}{|x-nL|^{\alpha+1}}  ,\hspace{0.5cm} \xi=\frac{x}{L} ,\\  \\
\displaystyle \hspace{3cm}K_L^{(\alpha)}(|x|) = \frac{\alpha !\sin{(\frac{\alpha\pi}{2})}}{\pi L^{\alpha+1}}\left\{-\frac{1}{|\xi|^{\alpha+1}}+
{\tilde \zeta}(\alpha +1,\xi) + {\tilde \zeta}(\alpha +1,-\xi) \right\}   \\  \\
\displaystyle \hspace{3cm}K_L^{(\alpha)}(|x|) = -\frac{\alpha !}{\pi} \lim_{\epsilon\rightarrow 0+} \Re\left\{\sum_{n=-\infty}^{\infty}
\frac{i^{\alpha+1}}{(x-nL+i\epsilon)^{\alpha+1}}\right\}  \\ \\
\hspace{0.5cm}\displaystyle =\frac{\alpha !}{\pi L^{\alpha+1}} \lim_{\epsilon\rightarrow 0+}
\Re \left\{\ i^{\alpha+1} \left(  \frac{1}{(\xi+i\epsilon)^{\alpha+1}}
-\zeta(\alpha+1,\xi+i\epsilon)-\zeta(\alpha+1,-\xi+i\epsilon) \right)\right\} .
\end{array}
\end{equation}

This kernel can be conceived as the explicit representation of the fractional Laplacian (Riesz fractional derivative) on the $L$-periodic string. 
The last relation is the distributional representation and is expressed by standard Hurwitz $\zeta$-functions denoted by $\zeta(..)$. The two variants of $\zeta$- functions which occur in above relation are defined by

\begin{equation}
\label{hurwitz}
{\tilde \zeta}(\beta,x)= \sum_{n=0}^{\infty}\frac{1}{|x+n|^{\beta}} ,\hspace{2cm} \zeta(\beta,x)=\sum_{n=0}^{\infty}
\frac{1}{(x+n)^{\beta}} ,\hspace{1.5cm} \Re\, \beta >1 .
\end{equation}
We see for $\alpha>0$ and $x\neq 0$ that the series in (\ref{perfrac}) are absolutely convergent as good as the power function integral  $\int_1^{\infty}\xi^{-\alpha-1}{\rm d}\xi$. For integer powers $\frac{\alpha}{2} \in {\bf N}$ the distributional representations (\ref{perfrac})$_{3,4}$ take the (distributional) forms of the (negative-semidefinite) 1D integer power Laplacian operators, namely

\begin{equation}
\label{integerperiodic}
\begin{array}{l}
\displaystyle K_L^{(\alpha=2m)}(|x|)= (-1)^{m+1}\frac{d^{2m}}{dx^{2m}} 
\sum_{n=-\infty}^{\infty} \lim_{\epsilon\rightarrow 0+}\frac{1}{\pi}\frac{\epsilon}{((x-nL)^2+\epsilon^2)} ,\hspace{1cm} \frac{\alpha}{2} = m\in {\bf N_0} ,\\ \\
\displaystyle \hspace{3cm} =
 (-1)^{m+1}\frac{d^{2m}}{dx^{2m}} \sum_{n=-\infty}^{\infty}\delta_{\infty}(x-nL) = -\left(-\frac{d^2}{dx^2}\right)^{\frac{\alpha}{2}=m}\delta_L(x) ,
\end{array}
\end{equation}
where $\delta_{\infty}(..)$ and $\delta_L$ indicate the Dirac's $\delta$-functions of the infinite and the $L$-periodic string, respectively.
We further observe in full correspondence to the discrete fractional Laplacian matrix, the necessary property that in the limit of an 
infinite string $\displaystyle \lim_{L\rightarrow\infty}  K_L^{(\alpha)}(|x|) = {\cal K}_{\infty}^{(\alpha)}(x) $ (\ref{perfrac}) recovers the expression of the standard 
1D infinite space fractional Laplacian kernel (\ref{inffyla}) 
known from the literture
(see for a further discussion \cite{michelJphysA,michelchaos} and references therein).

\section{Fractional Laplacian matrix on cubic lattices: towards fractional lattice dynamics}

In this section we deduce the $nD$ counterpart of the fractional Laplacian matrix introduced above. With that approach the fundamentals of 
`{\it fractional lattice dynamics}' can be deduced as a generalization of conventional lattice dynamics.

In this section our goal is to generalize the above 1D lattice approach to cubic periodic lattices in $n=1,2,3,..$ 
dimensions of the physical space where the 1D lattice case is contained. We assume the lattice contains $N=N_1..\times N_n$ lattice points, each covered by identical atoms with mass $\mu$. 
Each mass point is characterized by $\vec{p}=(p_1,p_2,..,p_n)$ ($p_j=0,..N_j-1$) and $n=1,2,3,..$ denotes the dimension of the physical space embedding the lattice. In order to define the lattice 
fractional Laplacian matrix, it is sufficient to consider
a {\it scalar} generalized displacement field $u_{\vec{p}}$ (one field degree of freedom) associated to each mass point $\vec{p}$ only. The physical nature of this scalar field can be any scalar field, 
such as for instance a one degree of freedom displacement field, an
electric potential or, in a stochastic context a probablitity density function (pdf) or in a fractional quantum mechanics context a Schr\"odinger wave function. This demonstrates the interdisciplinary
character of the present fractional lattice approach.

The fractional Laplacian matrix for general networks was only recently and to our knowledge for the first time introduced by Riascos and Mateos \cite{riascos-fracdyn,riascos-fracdiff} in the framework of fractional diffusion analysis 
on networks which include nD periodic lattices (nD tori) as special cases being subject of the present analysis. For cubic nD lattices the fractional Laplacian matrix can be written as 
\cite{michelJphysA,michelchaos,riascos-fracdyn,riascos-fracdiff}

\begin{equation}
\label{fraalphn}
\Delta_{\alpha,n} = -\mu \Omega_{\alpha,n}^2L_n^{\frac{\alpha}{2}} ,\hspace{1cm} L_n^{\frac{\alpha}{2}} = \left(2n {\hat 1}-A_n \right)^{\frac{\alpha}{2}} ,
\hspace{0.5cm} \alpha > 0 ,
\end{equation}
where ${\hat 1}$ denotes the identity matrix, $n$ indicates the dimension of the physical space and $2n$ indicates the connectivity, i.e. the number of next neighbors of a lattice point in the nD cubic lattice.
In (\ref{fraalphn}) we introduced the adjacency matrix $A_n$ which has for the cubic lattice with next neighbor connections the form

\begin{equation}
\label{cubicadj}
A_n = \sum_{j=1}^n (D_j+D_j^{\dagger}) ,
\end{equation}
where then $D_j$ and $D_j^{\dagger}=D_j^{-1}$ denote the next neighbor shift operators in the $j=1,..,n$-directions defined by $D_ju_{p_1,..p_j,..,p_n}= \vec{u}_{p_1,..p_j+1,..,p_n}$ 
and $D_j^{\dagger}\vec{u}_{p_1,..p_j,..,p_n}= u_{p_1,..p_j-1,..,p_n}$, i.e.
$D_j$ shifts the field associated to lattice point $\vec{p}=(..,p_j,..$ to the field associated with the adjacent lattice point in the positive $j$-direction $(..,p_{j+1},..)$, and the inverse (adjoint) shift operator 
$D_j^{\dagger}=D_j^{-1}$ to the adjacent lattice point in the negative $j$-direction $(..,p_{j-1},..)$. All matrices introduced in (\ref{fraalphn}) 
and (\ref{cubicadj})
are defined on the 
nD lattice being $N \times N$ matrices ($N=N_1 \times ..\times N_n$).
As in the case of 1D lattice the so defined fractional Laplacian matrix (\ref{fraalphn}) describes for non-integer 
powers $ \frac{\alpha}{2}$, $\notin {\bf N} $ nonlocal elastic interactions, whereas they are generated by the `local' next neighbor Born von Karman Laplacian which is in our definition up to a negative
dimension factor $-\mu\Omega_{2}$ equal to $L_n$. We therefore refer to $L_n$ as `generator matrix'. We emphasize that the sign convention of what 
we call `(fractional) Laplacian matrix' varies in the literature 
(e.g. by denoting the positive semidefinite matrix $L_n^{\frac{\alpha}{2}}$ as `fractional Laplacian matrix', this convention is chosen, e.g. in \cite{riascos-fracdyn,riascos-fracdiff}). 
We have chosen to refer to as `fractional Laplacian matrix'
the negative-semidefinite matrix
$-\mu\Omega_{\alpha}^2L_n^{\frac{\alpha}{2}}$ to be in accordance with the negative definiteness of continuum limit fractional Laplacian (\ref{integerperiodic}) containing as a special case $\frac{\alpha}{2}=1$ the negative semidefinite
conventional Laplacian $\frac{d^2}{dx^2}\delta_L(x-x')$).
For a discussion of some general properties of the fractional Laplacian (\ref{fraalphn}) well defined on general networks including $nD$ lattices, we refer to \cite{riascos-fracdyn,riascos-fracdiff}.
In the periodic and infinite lattice the shift operators are unitary. 
Assuming $N_j$-periodicity in each direction $j$, the fractional Laplacian matrix is defined by
the spectral properties of the $L_n$-matrix, namely by
\begin{equation}
\label{spectralfrac}
[L_n^{\frac{\alpha}{2}}]_{(\vec{p}-\vec{q})}=\frac{1}{N}
\sum_{\vec{\ell}} e^{i\vec{\kappa}_{\vec{\ell}}\cdot(\vec{p}-\vec{q})}\lambda_{\vec{\ell}}^{\frac{\alpha}{2}} ,
\hspace{0.5cm} \lambda_{\vec{\ell}} = \left(2n-2\sum_{j=1}^n\cos{(\kappa_{\ell_j})}\right) ,\,\alpha >0,
 \end{equation}
where we denoted $\sum_{\vec{\ell}}(..)=\sum_{\ell_1=0}^{N_1-1}(..)..\sum_{\ell_n=0}^{N_n-1}(..)$ and $\vec{\kappa}_{\vec{\ell}}=(\kappa_{\ell_1},..\kappa_{\ell_n}) $ denotes the Bloch wave vectors of the Brillouin zone where their components
can the values $\kappa_{\ell_j}=\frac{2\pi}{N_j}\ell_j$ ($\ell_j=0,..,N_j-1$). 
It can be seen that (\ref{spectralfrac}) has T\"oplitz structure depending only on $|p_1-q_1|,..,|p_j-q_j|,..|p_n-q_n|$).
For the infinite lattice when all $N_j\rightarrow\infty$ in (\ref{spectralfrac}), the summation over the
reciprocal lattice points assumes asymptotically the form of an integral 
$\frac{1}{N}\sum_{\vec{\ell}}g(\vec{\kappa}_{\ell}) \sim \frac{1}{(2\pi)^n}\int_{-\pi}^{\pi}..\int_{-\pi}^{\pi}{\rm d}\kappa_1..{\rm d}\kappa_n
g(\vec{\kappa})$, where the integration intervals $[-\pi,\pi]$ can be chosen instead of $[0,2\pi]$ for $2\pi$-periodic functions $g(\kappa_j)=g(\kappa_j+2\pi)$.
\vskip0.5cm
\begin{center}
\includegraphics[scale=0.32]{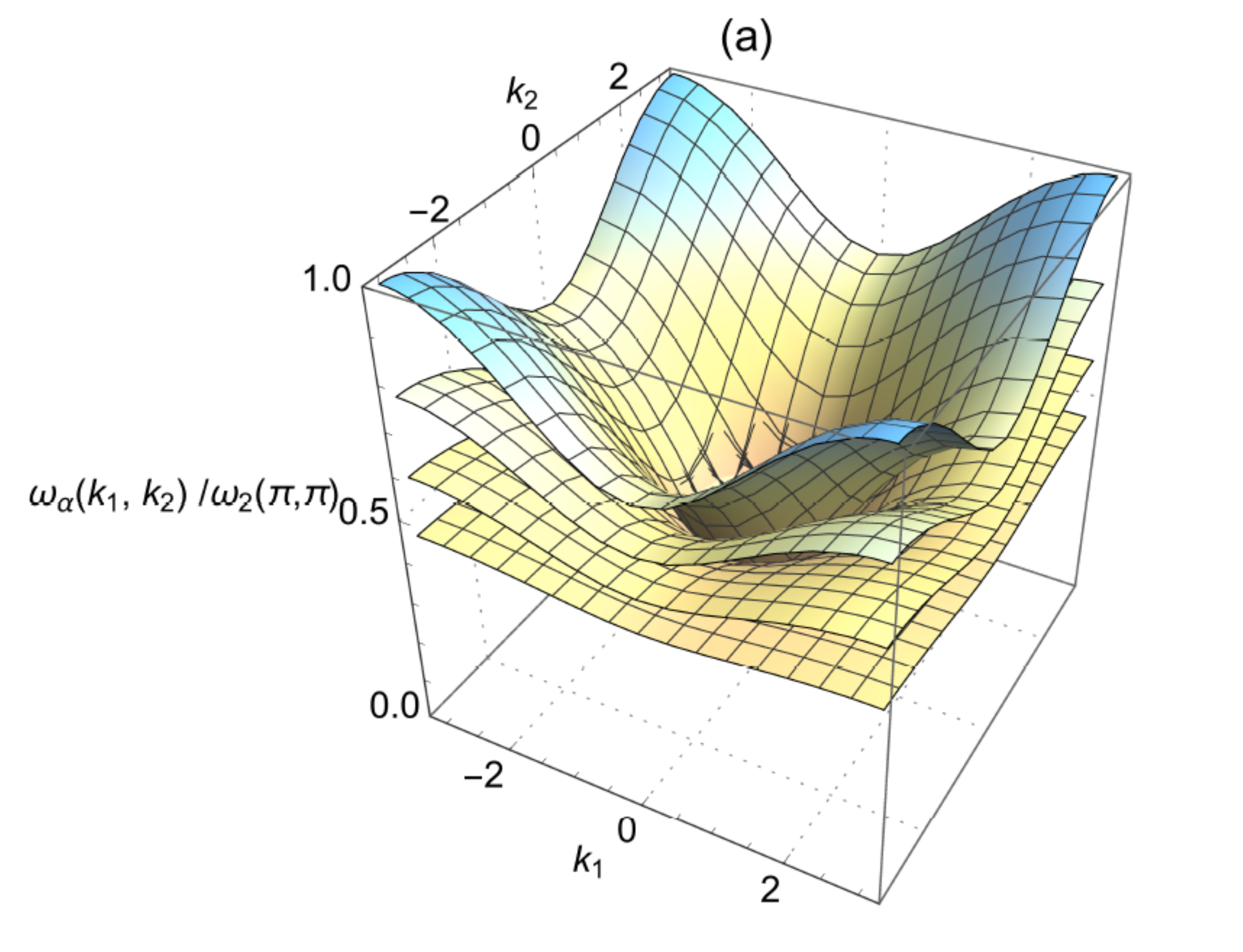}
\end{center}
\vskip-2cm
\begin{center}
\includegraphics[scale=0.32]{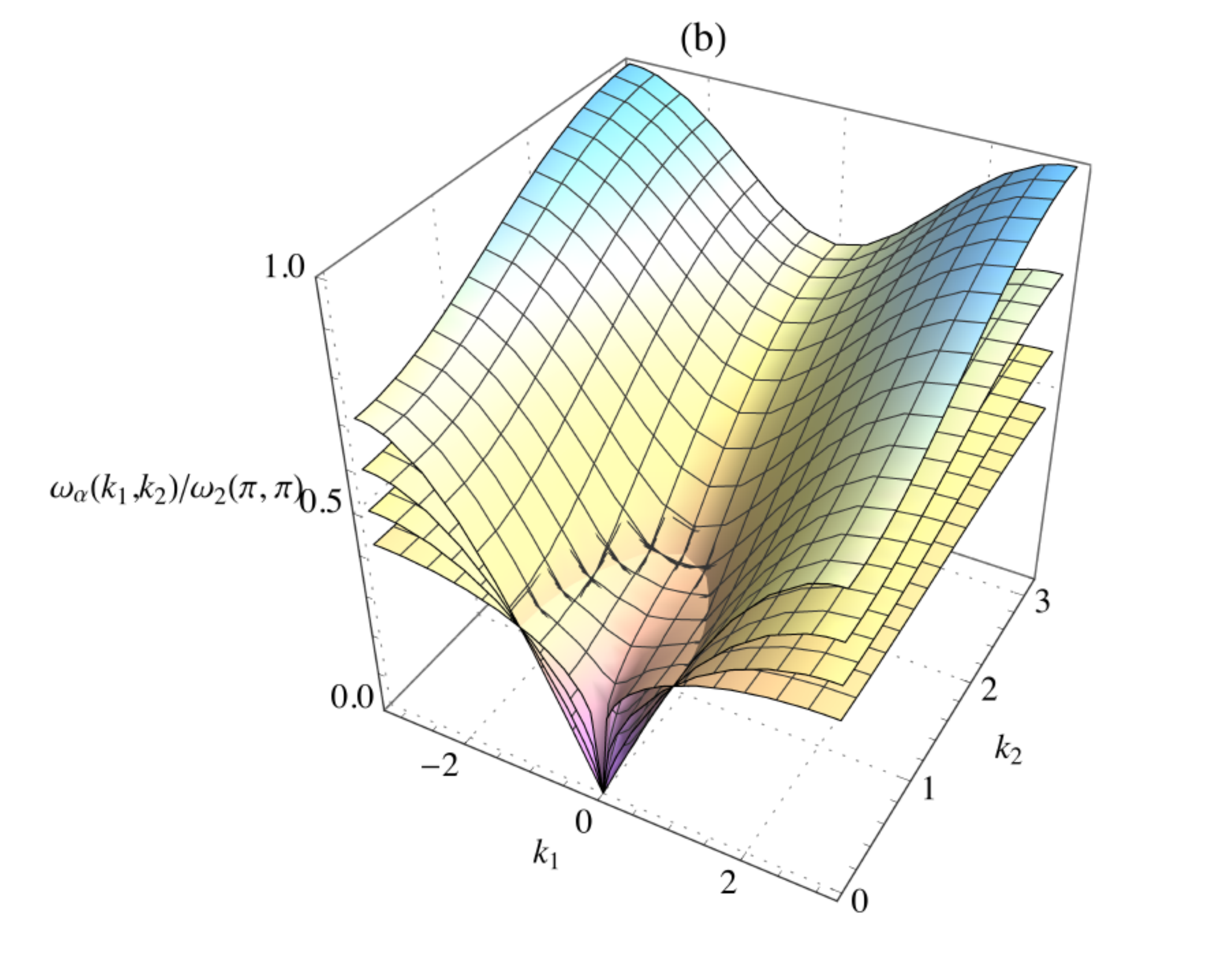}
\end{center}
\begin{center}
\includegraphics[scale=0.32]{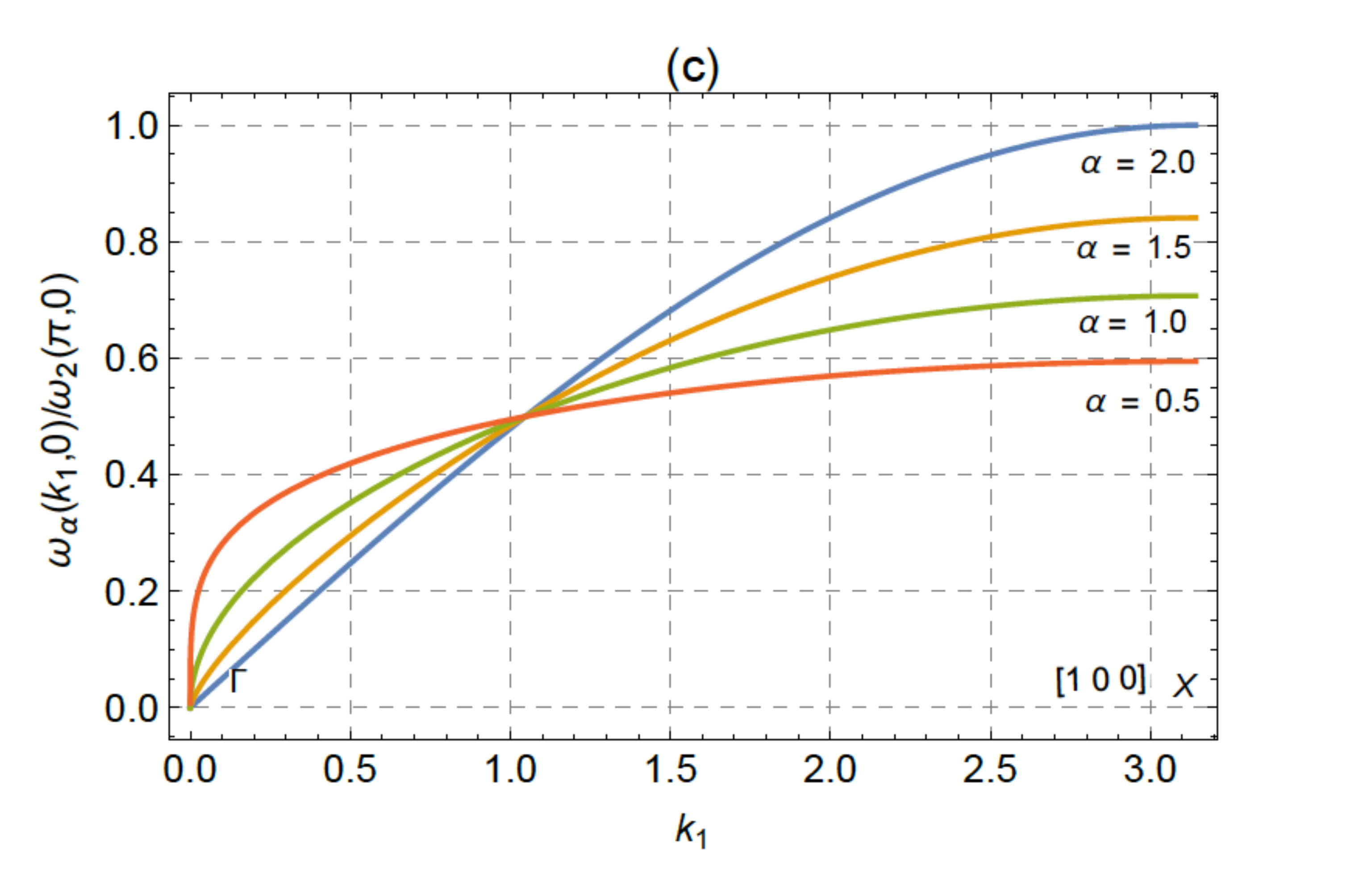}
\end{center}
\begin{center}
\includegraphics[scale=0.32]{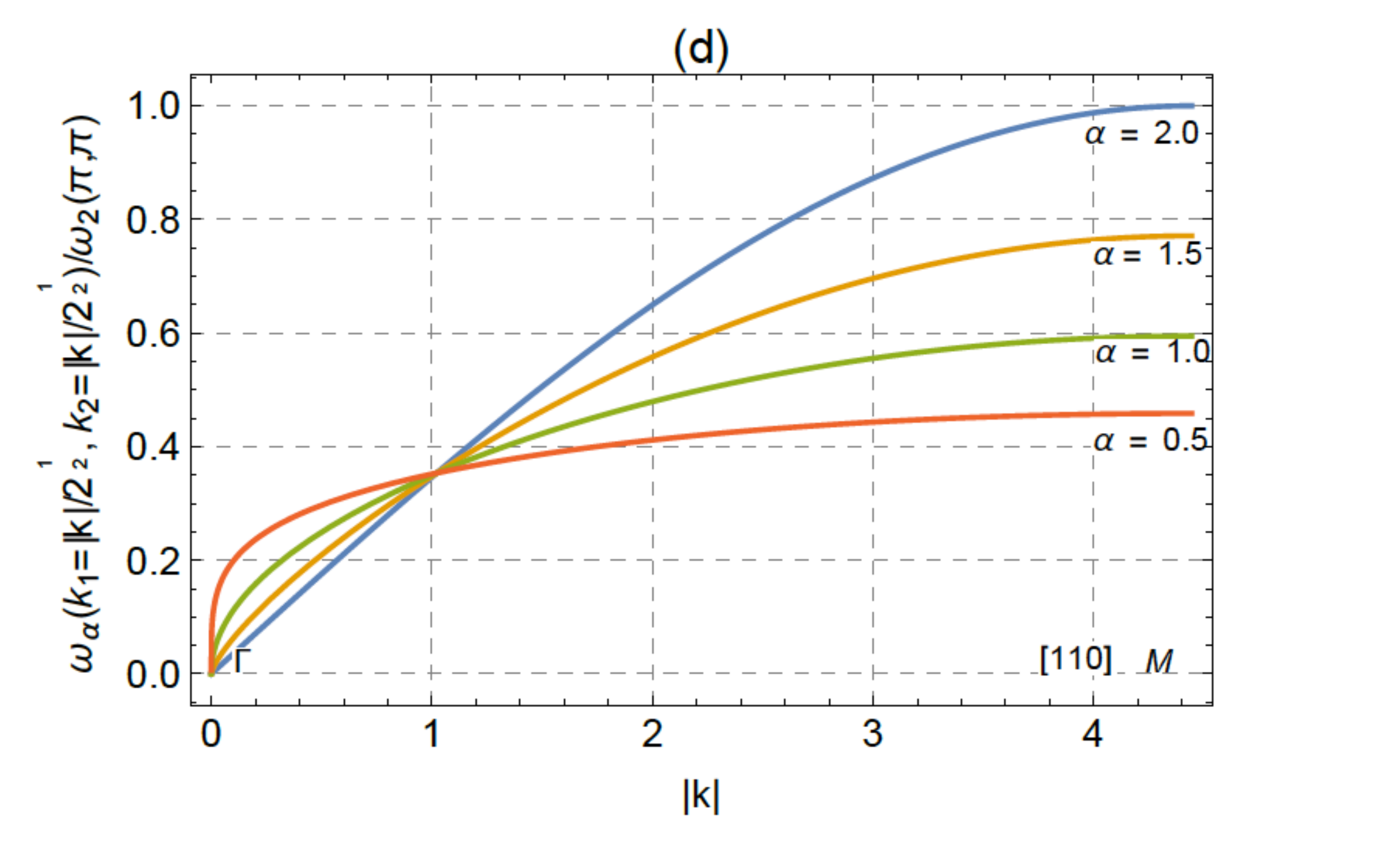}
\end{center}
{\footnotesize $\bold{Fig. 1}$ (a-b) 
Show the dispersion surfaces $\omega_{\alpha}(\kappa_1,\kappa_2)/\omega_{2}(\pi,\pi)= 
\lambda^{\frac{\alpha}{4}}(\kappa_1,\kappa_2)/2^{\frac{3}{2}} = 2^{\frac{\alpha-3}{2}} (sin^2(\kappa_1/2)+ sin^2(\kappa_2/2))^{\frac{\alpha}{4}} $ for the 2D cubic 
lattice ($n=2$) of (\ref{spectralfrac}) for four values of $\alpha$, while (c-d) illustrate cross-sections of these dispersion sheets 
with the planes (0 1 0) and (1 1 0), respectively.}
 \newline\newline
 For  $\alpha$ fixed, the circular frequency is given by $\omega_{\alpha}(\kappa_1,\kappa_2) = \lambda^{\frac{\alpha}{4}}$. 
 The linear frequency spectra $(a, b, d)$, for $n=2$, are normalized by the maximum frequency 
  $\omega_{\alpha=2}(\pi,\pi)= \lambda^{\frac{1}{2}}(\pi,\pi)= 2^{\frac{3}{2}} $ obtained for 
  a wave vector located in $(001)$ plane. It will be noted that the sheets cut at dimensionless frequency 
 $\omega_{\alpha}(\kappa_1,\kappa_2)/\omega_{\alpha=2}(\pi,\pi) \approx 0.351$ 
 and the dispersion relations of the classical next neighbor Born von Karman lattice are recovered (indicated by 
 $\omega_{\alpha}(\kappa_1,\kappa_2)/\omega_{\alpha=2}(\pi,\pi) \rightarrow 1$ for $\alpha=2$ and $\kappa_{1,2}\rightarrow \pi$,  
 $\omega_{\alpha}(\kappa_1,0)/\omega_{\alpha=2}(\pi,0) \rightarrow 1$ for $\alpha=2$ and $\kappa_1\rightarrow \pi$).
 When the value of $\alpha$ decreases, one observes  
 in agreement with another work \cite{michelJphysA}, namely a decrease of the maximum dimensionless frequency in end of the first Brillouin zone.

The goal is now to deduce a more convenient integral representation of (\ref{spectralfrac}). To this end we utilize the following observation: Let in the following ${\cal L}$ be a positive 
semidefinite\footnote{i.e. all eigenvalues $\lambda_{\ell}$ 
of this matrix are non-negative.} matrix and $\alpha >0$
 like (\ref{fraalphn}) with the spectral representation
\begin{equation}
 \label{spere}
{\cal L}=  \sum_{\vec{\ell}}\lambda_{\vec{\ell}} |\vec{\ell}><\vec{\ell}| ,\hspace{2cm} {\cal L}_{pq} = <p|{\cal L}q> ,
\end{equation}
where we have to put for the periodic nD lattice of (\ref{spectralfrac}) the Bloch-eigenvectors $<\vec{p}|\vec{\ell}> = N^{-\frac{1}{2}} e^{i\vec{\kappa}_{\vec{\ell}}\cdot\vec{p}}$.
Then it will be useful to define the matrix Dirac $\delta$-function by 
\begin{equation}
 \label{matrixdelta}
 \delta({\cal L}-\tau{\hat 1}) = \sum_{\vec{\ell}} |\vec{\ell}><\vec{\ell}|\, \delta(\tau-\lambda_{\vec{\ell}}) ,
\end{equation}
where $\tau$ is a scalar parameter and ${\hat 1}$ the identity matrix and $\delta(\tau-\lambda_{\vec{\ell}})$ 
the conventional scalar Dirac $\delta$-function. Then with the matrix $\delta$-function defined in
(\ref{matrixdelta}) we can write
\begin{equation}
 \label{matrixpower}
{\cal L}^{\frac{\alpha}{2}} = \int_{-\infty}^{\infty}\delta(L-\tau{\hat 1})|\tau|^{\frac{\alpha}{2}}{\rm d}\tau 
\end{equation}
and by utilizing $\delta(\tau-\lambda_{\vec{\ell}}) = \frac{1}{(2\pi)}\int_{-\infty}^{\infty}e^{ik(\tau-\lambda_{\vec{\ell}})}{\rm d}k$ 
together with the kernel $-{\cal D}_{\frac{\alpha}{2}}$ of the 1D fractional Laplacian (Riesz fractional derivative) of order $\frac{\alpha}{2}$ in its distributional form \cite{michel-et-al-14}
\begin{equation}
 \label{distrifoufracder}
 \begin{array}{l}
 {\cal D}_{\frac{\alpha}{2}}(k-\xi) = \displaystyle \left(-\frac{d^2}{dk^2}\right)^{\frac{\alpha}{4}}\delta(k-\xi) =:
\frac{1}{(2\pi)}\int_{-\infty}^{\infty}e^{i(k-\xi)\tau}|\tau|^{\frac{\alpha}{2}}{\rm d}\tau =  \\ \\
\displaystyle \lim_{\epsilon\rightarrow 0+}\frac{1}{\pi}\Re \int_0^{\infty}e^{-\tau(\epsilon-i(k-\xi))}|\tau|^{\frac{\alpha}{2}}{\rm d}\tau = 
 \lim_{\epsilon\rightarrow 0+} \Re \frac{\Gamma(\frac{\alpha}{2}+1)}{\pi(\epsilon-i(k-\xi))^{\frac{\alpha}{2}}} .
\end{array}
\end{equation}
Then we can write for the matrix power function (\ref{matrixpower}) the representation
\begin{equation}
 \label{repre}
 {\cal L}^{\frac{\alpha}{2}} = \int_{-\infty}^{\infty} e^{ik{\cal L}}{\cal D}_{\frac{\alpha}{2}}(k) {\rm d}k ,
\end{equation}
where the exponential $e^{ik{\cal L}}$ of the matrix ${\cal L}=L_n$ can be determined more easily for the generator $L_n = \sum_{j=1}^n L_j$  ($L_j=2-D_j-D_j^{\dagger}$) being the sum of the 1D generator matrices of the $N_j$-periodic
1D lattices and having therefore the eigenvalues $\lambda(\ell_j)=2-2\cos{\kappa_{\ell_j}}$ and as a consequence having a Cartesian product space spanned by the periodic Bloch eigenvectors 
$\frac{e^{i{\vec \kappa}_{\vec{\ell}}\cdot \vec{p}}}{\sqrt{N}} =\prod_{j=1}^n \frac{e^{ip_j\kappa_{\ell_j}}}{\sqrt{N_j}} $.
The matrix elements of the spectral representation of the exponential of $L_n$ can hence be written as
\begin{equation}
 \label{expolapl}
 [e^{i\xi L_n}]_{\vec{p}-\vec{q}} = \sum_{\vec{\ell}} \frac{e^{i{\vec \kappa}_{\vec{\ell}}\cdot (\vec{p}-\vec{q})}}{N}e^{i\xi \lambda_{\vec{\ell}}} = 
 \prod_{j=1}^n \sum_{\ell_j=1}^{N_j-1} \frac{e^{i(p_j-q_j)\kappa_{\ell_j}}}{N_j}e^{i2k(1-\cos{\kappa_{\ell_j}})} .
\end{equation}

\noindent {\bf Infinite nD lattice}\\ 
In the limiting case of an infinite nD lattice when all $N_j\rightarrow\infty$ we can write by using $\frac{1}{N}\sum_{\vec{\ell}}g({\vec \kappa}_{\vec{\ell}}) \sim \frac{1}{(2\pi)^n} 
\int_{-\pi}^{\pi}..\int_{-\pi}^{\pi}{\rm d}\kappa_1..{\rm d}\kappa_n g({\vec \kappa}) $ to arrive at

\begin{equation}
 \label{expolaplinfini}
 [e^{i\xi L_n}]_{\vec{p}-\vec{q}} = [e^{i\xi L_n}]_{|p_1-q_1|,..,|p_n-q_n|} =
 \prod_{j=1}^n \frac{1}{(2\pi)}\int_{-\pi}^{\pi} e^{i(p_j-q_j)\kappa}e^{i2\xi (1-\cos{\kappa})}{\rm d}\kappa .
\end{equation}
Taking into account the definition of the modified Bessel functions of the first kind $I_p(z)= \frac{1}{\pi}\int_0^{\pi}e^{z\cos{\varphi}}\cos{p\varphi}{\rm d}\varphi$ where $p={\bf N}_0$ 
denotes non-negative integers \cite{abramo}, we can write the exponential matrix (\ref{expolaplinfini}) in the form 

\begin{equation}
 \label{expobessel}
 [e^{i\xi L_n}]_{|p_1-q_1|,..,|p_n-q_n|} = e^{i 2n\xi} \prod_{j=1}^nI_{|p_j-q_j|}(-2i\xi) .
\end{equation}
Applying now the matrix relation (\ref{matrixpower}) and plugging in the exponential (\ref{expobessel}) yields an integral representation of the (negative semidefinite) fractional Laplacian matrix (\ref{fraalphn}) in terms of a product of
modified Bessel functions of the first kind, namely
\begin{equation}
 \label{fractionallapla}
 \begin{array}{l}
 \displaystyle [\Delta_{\alpha,n}]_{|p_1-q_1|,..,|p_n-q_n|}  = -\mu \Omega_{\alpha,n}^2L^{\frac{\alpha}{2}}_{|p_1-q_1|,..,|p_n-q_n|}   \\ \\
 \hspace{1cm} \displaystyle = -\mu \Omega_{\alpha,n}^2 \int_{-\infty}^{\infty} {\rm d}\xi \, e^{i 2n\xi} {\cal D}_{\frac{\alpha}{2}}(\xi) \prod_{j=1}^nI_{|p_j-q_j|}(-2i\xi) ,
 \end{array}
\end{equation}
with $-{\cal D}_{\frac{\alpha}{2}}(\xi)$ indicating the Riesz fractional derivative kernel of (\ref{distrifoufracder}). 

\noindent {\bf Asymptotic behavior}\\
Introducing the new vector valued integration variable ${\vec{\xi}} = \vec{\kappa} p $ ($\xi_j=p\kappa_j , \forall j=1,..,n$) we can write for the infinite lattice integral of (\ref{spectralfrac}) 
by utilizing spherical polar coordinates $ \vec{p} =p \vec{e}_{\vec{p}} $  ($ \vec{e}_{\vec{p}}\cdot \vec{e}_{\vec{p}}=1$, $p^2=\sum_j^n p_j^2$) 
\begin{equation}
 \label{recrit}
 L_n^{\frac{\alpha}{2}}({\bf p}) =  \frac{1}{(2\pi)^n} \int_{-\pi p}^{\pi p}..\int_{-\pi p}^{\pi p}\frac{{\rm d}\xi_1..{\rm d}\xi_n}{p^n}
 \left(4\sum_{j=1}^n\sin^2{\frac{\xi_j}{2p}} \right)^{\frac{\alpha}{2}}\cos{(\vec{\xi}\cdot \vec{e}_{\vec{p}})} .
\end{equation}
The dominating term for $p>>1$ becomes 

\begin{equation}
 \label{recrit2}
 L_n^{\frac{\alpha}{2}}({\bf p}) \approx \frac{1}{p^{n+\alpha}} \frac{1}{(2\pi)^n} \int_{-\infty}^{\infty}..\int_{-\infty}^{\infty}{\rm d}\xi_1..{\rm d}\xi_n
\left( \sum_{j=1}^n \xi_j^2\right)^{\frac{\alpha}{2}}\cos{(\vec{\xi}\cdot \vec{e}_{\vec{p}} )} ,
\end{equation}
having the form
\begin{equation}
 \label{recrit3}
 L_n^{\frac{\alpha}{2}}(\vec{p})_{p>>1} \approx - \frac{C_{n,\alpha}}{p^{n+\alpha}} ,
\end{equation}
where the positive normalization constant is obtained explicitly as $C_{n,\alpha}=\frac{2^{\alpha-1}\alpha\Gamma(\frac{\alpha+n}{2})}{\pi^{\frac{n}{2}}\Gamma(1-\frac{\alpha}{2})}$, e.g. \cite{michel-et-al-13,michel-et-al-14}. 
We can identify the asymptotic representation (\ref{recrit2}), (\ref{recrit3}) with the kernel of Riesz fractional derivative (fractional Laplacian) 
of the nD infinite space. For a more detailed discussion of properties we refer to \cite{michel-et-al-13,michel-et-al-14}.

\section{Conclusions}

We have introduced a fractional lattice dynamics approach which defines exact expressions for fractional lattice Laplacian matrices on nD periodic and infinite lattices. These fractional Laplacian matrices
have all `good' properties of Laplacian matrices (translational invariance and negative semidefiniteness). The formulation of our approach is fully consistent with the fractional network approach of Riascos and 
Mateos \cite{riascos-fracdyn,riascos-fracdiff}.
In the infinite space and periodic lattice continuum limits these fractional Laplacian matrices take the representations of the well known respective Riesz fractional derivative kernels, i.e. 
the convolutional kernels of the (continuous) fractional Laplacians. The approach allows to model `anomalous diffusion' phenomena on lattices with fractional transport phenomena including asymptotic emergence of L\'evy flights.
In such a lattice diffusional model, the conventional Laplacian matrix is generalized by its fractional power law matrix function counterpart. 

As a general framework, the present fractional lattice Laplacian appears to be fundamental in various physical 
contexts, especially as a point of departure for a newly emerging generalization of Lattice Dynamics to `{\it Fractional Lattice Dynamics}'.

\section*{Acknowledgements}
{\it  Fruitful discussions with G.A. Maugin, A. Porubov are greatfully acknowledged.}

\bigskip

\address{{\rm\bf  Thomas M. Michelitsch, Bernard A. Collet} \\ Sorbonne Universit\'es, Universit\'e Pierre et Marie Curie (Paris 6), Institut Jean le Rond d'Alembert, CNRS UMR 7190, 4 place Jussieu, 75252 Paris cedex 05, France}

\address{{\rm\bf Alejandro P. Riascos}\\ Instituto de F\'{i}sica, Universidad Nacional Aut\'{o}noma de M\'{e}xico, Apartado Postal 20-364, 01000 M\'{e}xico, D.F., M\'{e}xico}

\address{{\rm \bf Andrzej F. Nowakowski, Franck C.G.A. Nicolleau}\\ Department of Mechanical Engineering, The University of Sheffield,
Sir Frederick Mappin Building,
Mappin Street,
Sheffield S1 3JD, United Kingdom}

\end{document}